\documentclass{eas}
\usepackage{graphicx}
\usepackage{graphicx}
\usepackage{latexsym}
\usepackage{amsmath}
\usepackage{amsfonts} 
\usepackage{amssymb}
\usepackage[latin1]{inputenc}
\def\beq{\begin{equation}}
\def\eeq{\end{equation}}
\def\rmd{{\rm d}}
\def\rmD{{\rm D}}
\def\rightcontract{\mathop{\hbox{\vrule width0.5pt height6pt%
  \vrule height0.5pt width6pt}}}
\title{Strains and Jets in black hole fields}
\begin{document}%
\author{D. Bini}\address{{Istituto per le Applicazioni del Calcolo {\it M. Picone}}
{ CNR  Rome, Italy}}
\author{ F. de Felice}\address{Dipartimento di Fisica {\it G. Galilei}, Universit\`a di Padova
 { INFN, Sezione di Padova, Italy}}
\author{ A. Geralico}\address{{Dipartimento di Fisica and ICRA}
{Universit\'a di Roma ``La Sapienza,'' Roma, Italy}}
\begin{abstract}
We study the behaviour of an initially spherical bunch of particles
emitted along trajectories parallel to the symmetry axis of a Kerr black hole. We show that, under suitable conditions, curvature and inertial strains compete to generate jet-like structures.
\end{abstract}
\maketitle
\section{Introduction}
In recent papers, Bini, de Felice and Geralico (2006, 2007) considered  how the relative deviations among the particles of a given bunch depend on the geometric properties of the  frame adapted to a fiducial observer. Here we outline the main steps of their analysis which 
revealed how spacetime curvature and inertial strains compete to generate jet-like structures. 

\section{The relative deviation equation}

Consider a bunch of test particles, i.e. a congruence ${\mathcal C}_U$ of timelike world lines with unit tangent vector
$U$ ($U\cdot U=-1$)  parametrized by the proper time $\tau_U$. In general, the lines of the congruence ${\mathcal C}_U$  are accelerated with  acceleration   
$a(U)=\nabla_U U$.
Let ${\mathcal C}_*$ be  a fixed world line of the congruence  which we consider as that of the \lq\lq fiducial observer." 
 The separation  between the line ${\mathcal C}_*$ and a line of ${\mathcal C}_U$ is represented by a connecting vector $Y$, i.e. a vector undergoing Lie transport along $U$, namely $\pounds_U Y=0$. 
The latter equation becomes:
\beq
\dot Y\equiv\frac{\rmD Y}{\rmd \tau_U }=\nabla_Y U= -(Y\cdot U)a(U)-K(U)\rightcontract Y\ ,
\eeq
where the kinematical tensor $K(U)$ is defined  as  $K(U)=\omega(U)-\theta(U)$. Here 
 the antisymmetric tensor $\omega(U)^\beta{}_\alpha$ represents the vorticity of the congruence ${\mathcal C}_U$  and the symmetric tensor $\theta(U)^\beta{}_\alpha$ represents the expansion.
The symbol $\rightcontract $ stands for right-contraction operation among tensors.  
The covariant derivative along $U$ of both sides of the Lie transport equation  gives rise to the \lq\lq relative deviation equation" which, once projected on a  given spatial triad $E_a$, reads:
\beq
\label{eq:secder3bis}
\ddot Y^a +{\mathcal K}_{(U,E)}{}^a{}_b  Y^b=0\ ,  
\eeq
where ${\mathcal K}_{(U,E)}{}^a{}_b=[T_{({\rm fw},U,E)}-S(U)+{\mathcal E}(U)]^a{}_b$. 
Here ${\mathcal E}(U)$ is the electric part of the Riemann tensor relative  to $U$ namely $
{\mathcal E}(U)^\alpha{}_\gamma =R^\alpha{}_{\beta\gamma\delta}U^\beta U^\delta$. 
The strain tensor $S$ is defined as
\beq
\label{eq:strainsdef}
S(U)_{ab}=\nabla(U)_b a(U)_a+a(U)_aa(U)_b\ ;
\eeq
$S(U)$ depends only on the congruence ${\cal C}_U$ and not on the chosen spatial triad $E_a$, while the tensor $T$, given by:  
\begin{eqnarray}
\label{Tdef}
&& T_{({\rm fw},U,E)}{}^a{}_b=\delta^a_b \omega_{({\rm fw},U,E)}^2-\omega_{({\rm fw},U,E)}^a \omega_{({\rm fw},U,E)}{}_b\nonumber \\
&& \quad -\epsilon^a{}_{bf}\dot \omega_{({\rm fw},U,E)}^f -2\epsilon^a{}_{fc}\omega_{({\rm fw},U,E)}^f K(U)^c{}_b\ , 
\end{eqnarray}
as derived in detail in Bini et al. (2006), characterizes the given frame relative to a Fermi-Walker frame along the congruence. 
We like to  stress here that the components of the 
deviation vector should be restricted to the reference observer's world line; they may behave differently according  to the  reference frame set up for their measurements.

\section{Axial observers in Kerr spacetime}

Consider Kerr spacetime, whose line element in Kerr-Schild coordinates $(t,x^1=x,x^2=y,x^3=z)$ is given by
\begin{eqnarray}
\label{metr_gen}
\rmd s^2 &=& g_{\alpha\beta}\rmd x^\alpha\rmd x^\beta\equiv (\eta_{\alpha\beta}+2Hk_\alpha k_\beta)\rmd x^\alpha\rmd x^\beta\ , 
\end{eqnarray}
where 
$H=\mathcal{M}r^3/(r^4+a^2z^2)$, 
$\eta_{\alpha\beta}$ is the flat spacetime metric and
\beq
k_\alpha\rmd x^\alpha=-\rmd t-\frac{(rx+ay)\rmd x+(ry-ax)\rmd y}{r^2+a^2}-\frac{z}{r}\rmd z\ ,
\eeq
with $r$ implicitly defined by
\beq
\frac{x^2+y^2}{r^2+a^2}+\frac{z^2}{r^2}=1\ .
\eeq
Here $\mathcal{M}$ and $a$ are the total mass and specific angular momentum characterizing the spacetime. 
Consider now a set of particles moving along the $z$-direction, on and nearby the axis of symmetry, with four velocity 
\beq
\label{boost1}
U=\gamma[m+\nu E(m)_{\hat 3}]\ , \qquad \gamma=(1-\nu^2)^{-1/2}\ ,
\eeq
where $m=\sqrt{(z^2+a^2)/\Delta_z}\partial_t$, $E(m)_{\hat 3}=\sqrt{\Delta_z/(z^2+a^2)}[\partial_z+(2{\mathcal M}z/\Delta_z)\partial_t]$, $\Delta_z=z^2-2{\mathcal M}z+a^2$ and 
where the instantaneous linear velocity $\nu=\nu(z)$, relative to the local static observers, is function of $z$.
These particles are accelerated except perhaps those which move strictly on the axis of symmetry.
Their history forms a congruence ${\mathcal C}_U$ of $\infty^2$ world lines and each of them can be parametrized by the pair $(x,y)$ of the spatial coordinates.
A frame adapted to this kind of orbits was found explitely in 
Bini at al. (2007) and termed $\{E(U)_{\hat a}\}$.  
The congruence is accelerated with non-vanishing components along the three spatial directions $E(U)_{\hat a}$. It is now possible to study the components of the vector $Y$ connecting a specific curve  of the congruence which we fix 
along the rotation axis with $x=0=y$ with tangent vector $\tilde U$ and nearby  world lines of the same congruence.
The reference world line is accelerated along the direction $E(\tilde U)_{\hat 3}$, i.e. $a(\tilde U)=a(\tilde U)^{\hat 3}E(\tilde U)_{\hat 3}$, with
\begin{eqnarray}
\label{acc}
 a(\tilde U)^{\hat 3}
&=&\frac{\rmd}{\rmd z}\left(\frac{\gamma}{\sqrt{\tilde g_{33}}}\right)\ ,
\end{eqnarray}
where $\tilde g_{33}=(z^2+a^2)/\Delta_z$ denotes the restriction on the axis of that metric coefficient.

\section{Tidal forces, frame induced deformations and strains}

In order to determine the deviations measured by the \lq\lq fiducial observer" $\tilde U$ with respect to the chosen  frame
we need to evaluate the kinematical fields of the congruence, namely  the acceleration $a(U)$, the vorticity $\omega(U)$ and the expansion $\theta(U)$), the electric part of the Weyl tensor ${\mathcal E}(U)$, the strain tensor $S(U)$ and the characterization of the spatial triad $\{E(U)_{\hat a}\}$ with respect to a Fermi-Walker frame given by the tensor $T_{({\rm fw},U,E)}$.  All these quantities are  calculated on the axis of symmetry so we can drop the tilde ($\, \tilde {}\, $); a lengthy calculation leads to the following results.
\\
\noindent
i) The only nonvanishing components of the kinematical tensors $\omega(U)$ and $\theta(U)$ 
 are given by
\beq 
\label{variousqua_rad}
\theta(U)_{\hat 3 \hat 3}
=\frac{\rmd}{\rmd z}\left(\frac{\gamma\nu}{\sqrt{g_{33}}}\right)\ , \ 
\omega(U)_{\hat 1 \hat 2}=\gamma(1+\nu)\frac{2a\mathcal{M}z}{\sqrt{\Delta_z}(z^2+a^2)^{3/2}}\ . 
\eeq 

\noindent
ii) The vorticity vector becomes $\omega_{({\rm fw},U,E)} =\omega_{({\rm fw},U,E)}{}_{\hat 3}E(U)_{\hat 3}$ 
with $\omega_{({\rm fw},U,E)}{}_{\hat 3}=\omega(U)_{\hat 1 \hat 2}$. 
\\
\noindent
iii) The electric part of the Weyl tensor is a diagonal matrix with respect to the adapted frame 
$\{E(U)_{\hat a}\}$, namely
\beq
\label{elermn}
{\mathcal E}(U)=\mathcal{M}z\frac{z^2-3a^2}{(z^2+a^2)^3}\,{\rm diag}\left[1,1,-2\right]\ .
\eeq

\noindent
iv) The only nonvanishing components of the tensor $T_{({\rm fw},U,E)}$ turn out to be 
\begin{eqnarray}
\label{variousqua_rad2}
&&T_{({\rm fw},U,E)}{}_{\hat 1 \hat 1}=T_{({\rm fw},U,E)}{}_{\hat 2 \hat 2}=-\omega_{({\rm fw},U,E)}{}_{\hat 3}^2\ , \nonumber\\
&&T_{({\rm fw},U,E)}{}_{\hat 1 \hat 2}=-T_{({\rm fw},U,E)}{}_{\hat 2 \hat 1}=-\dot\omega_{({\rm fw},U,E)}{}_{\hat 3}\nonumber\\
&&=\frac{\gamma\nu}{\sqrt{g_{33}}}\frac{\rmd }{\rmd z}\omega_{({\rm fw},U,E)}{}_{\hat 3}\ .
\end{eqnarray}

\noindent
v) The non-zero components of the strain tensor  $S(U){}^{\hat a}{}_{\hat b}$ can be written as  
\begin{eqnarray}
\label{strainsgen}
S(U){}_{\hat 1 \hat 1}&=&S(U){}_{\hat 2 \hat 2}=-\omega_{({\rm fw},U,E)}{}_{\hat 3}^2+\mathcal{M}z\frac{z^2-3a^2}{(z^2+a^2)^3}\ , \nonumber\\
S(U){}_{\hat 1 \hat 2}&=&-S(U){}_{\hat 2 \hat 1}=-\dot\omega_{({\rm fw},U,E)}{}_{\hat 3}\ , \nonumber\\
S(U){}_{\hat 3 \hat 3}&=&\frac{\gamma}{\sqrt{g_{33}}}\frac{\rmd }{\rmd z}[a(U)^{\hat 3}]+[a(U)^{\hat 3}]^2\ .
\end{eqnarray}

From Eqs. (\ref{elermn})--(\ref{strainsgen}) it follows that the deviation matrix ${\mathcal K}_{(U,E)}$ 
has only the non-zero component
\beq
\label{Kab}
{\mathcal K}_{(U,E)}{}_{\hat 3 \hat 3}=-\frac{\gamma}{\sqrt{g_{33}}}\frac{\rmd [a(U)^{\hat 3}]}{\rmd z}-[a(U)^{\hat 3}]^2+{\mathcal E}(U)_{\hat 3 \hat 3}\ ;
\eeq
hence, as expected, the particles emitted nearby the axis in the $z$-direction will be relatively accelerated in the $z$-direction only.
The spatial components of the connecting vector $Y$ are then obtained by integrating the deviation equation (\ref{eq:secder3bis}) which now reads
\begin{eqnarray}
\label{eq2ordrad}
\ddot Y^{\hat 1}=0\ , \quad 
\ddot Y^{\hat 2}=0\ , \quad
\ddot Y^{\hat 3}=-{\mathcal K}_{(U,E)}{}_{\hat 3 \hat 3}Y^{\hat 3}\ .
\end{eqnarray}
Equations  (\ref{eq2ordrad}) can be analytically integrated to give
\beq
\label{solY}
Y^{\hat 1}=Y^{\hat 1}_0,\qquad Y^{\hat 2}=Y^{\hat 2}_0, \qquad Y^{\hat 3}=C\frac{\gamma\nu}{\sqrt{g_{33}}}\ ,
\eeq
where $C$ is a constant.
This result shows that the conditions imposed on the particles of the bunch to move parallel to the axis of rotation is assured by a suitable balancing 
among the gravitoelectric (curvature) tensor (\ref{elermn}), the inertial tensor (\ref{variousqua_rad2}) and the strain tensor (\ref{strainsgen}).

\section{Reference world line uniformly accelerated}

\begin{figure} 
\typeout{*** EPS figure 1}
\begin{center}
\includegraphics[scale=0.45]{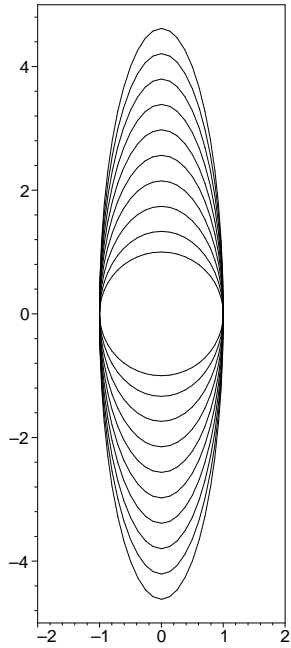}
\end{center}
\caption{An initially circular bunch of particles on the $y^{\hat 1}$-$y^{\hat 3}$ plane emitted along the $z$-axis at $z_0/\mathcal{M}=2$ is shown to spread for the choice of parameters $a/\mathcal{M}=0.5$, $\bar \kappa=1.5$ and ${\mathcal M}\, A=0.3$ as an example.  
The curves correspond to increasing values of the coordinate $z/\mathcal{M}=[2,4,6,8,10,12,14,16,18,20]$.
}
\label{fig:1}
\end{figure}

\begin{figure} 
\typeout{*** EPS figure 2}
\begin{center}
\includegraphics[scale=0.45]{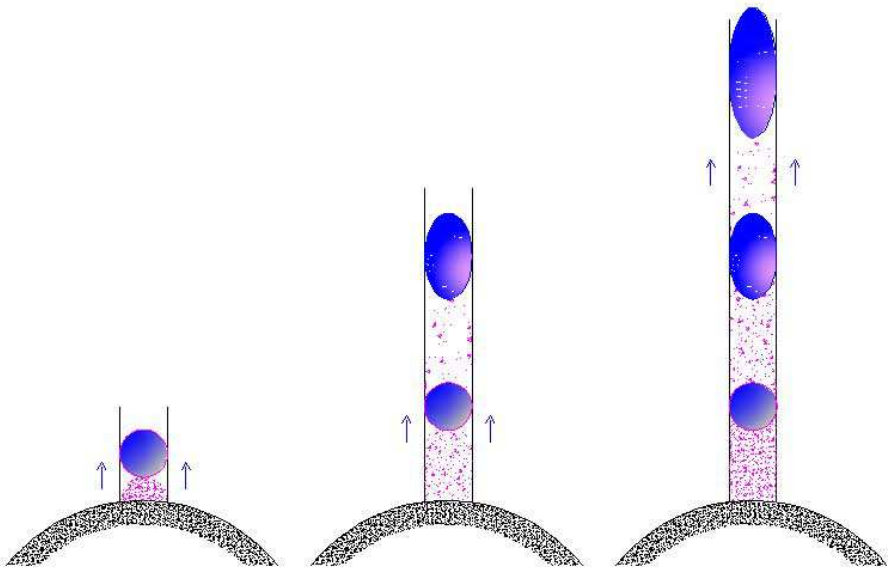}
\end{center}
\caption{The qualitative behaviour of an initially spherical bunch of particles moving out along the  axis of a rotating black hole.}
\label{fig:2}
\end{figure}

Let us now consider the case of the reference world line with tangent vector $U$ constantly accelerated, namely with  $a(U)^{\hat 3}=A=\,\,$const.
The instantaneous linear velocity relative to a local static observer is given by 
\beq
\label{eq:nua}
\nu_A=\left[1-\frac{1}{g_{33}[\bar\kappa+A(z-z_0)]^2}\right]^{1/2}\ ,
\eeq 
 where the positive value has been selected for $\nu_A$ in order to consider outflows.
 $\bar\kappa$ is a constant which  acquires the meaning of the conserved (Killing) energy of the particle
if the latter is not accelerated.
The solution of  equation (\ref{solY}) 
is given by
\beq
\label{solYaccrad}
Y^{\hat 1}=Y^{\hat 1}_0\ , \qquad 
Y^{\hat 2}=Y^{\hat 2}_0\ , \qquad 
Y^{\hat 3}=Y^{\hat 3}_0\frac{\gamma_A\nu_A}{\sqrt{g_{33}}\bar\kappa \nu_{A}^0}\ ,
\eeq
where $\nu_A^0=\nu_A(z_0)$.
Figure \ref{fig:1} shows the behaviour of an initially spherical bunch of particles in the $Y^{\hat 1}$-$Y^{\hat 3}$ plane for increasing values of the coordinate $z$. We clearly see a 
stretching along the $z$ axis leading to a collimated axial outflow of matter,  clearly suggestive of an  \lq\lq astrophysical jet."  Figure \ref{fig:2} shows qualitatively the behaviour of an initially spherical bunch of particles. 
Note that in this case the acceleration acts contrarily to the curvature tidal effect; indeed, 
$S(U)$, $T_{({\rm fw},U,E)}$ and ${\mathcal E}(U)$ act in competition leading to a quite unexpected result.
It is easy to show that 
the stretching  shown in Figs. \ref{fig:1} and \ref{fig:2} for uniformly accelerated outgoing particles persists also with a general acceleration. Moreover this behaviour does not
depend of the acceleration mechanism itself. 
Observed jets (see Figs. 3 and 4 as an example) appear to contain spherical bunches of particles emerging from the central black hole. 
\bibliographystyle{aipprocl}

\begin{figure} 
\typeout{*** EPS figure 3}
\begin{center}
\includegraphics[scale=0.65]{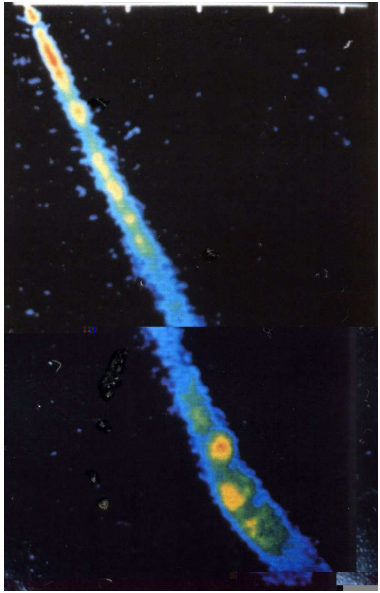}
\caption{The jet from the Galaxy M87 (Credit:J.A. Biretta et al. Hubble Heritage Team (STScl/AURA), NASA)
We can see almost spherical bunches of particles along the jet.}
\end{center}
\label{fig:3}
\end{figure}

\begin{figure} 
\typeout{*** EPS figure 4}
\begin{center}
\includegraphics[scale=0.45]{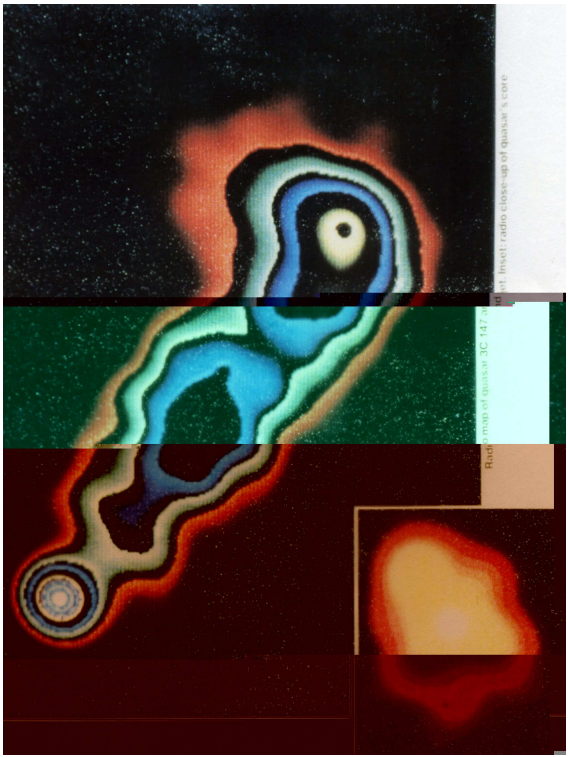}
\end{center}
\caption{Details of the jet from M87 close to the central black hole (Credit:J.A. Biretta et al. Hubble Heritage Team (STScl/AURA), NASA). Here we see a spherical bunch of particles apparently emerging from nearby the hole.}
\label{fig:4}
\end{figure}

\end{document}